\documentclass[a4paper,fleqn]{cas-dc}
\usepackage{lipsum}
 \usepackage{placeins}
\usepackage{shortcuts}
\usepackage{fontawesome}
\usepackage[labelsep=newline,justification=centering]{caption} 
\usepackage{mwe}
\usepackage{wasysym}
\usepackage{svg}
\usepackage[numbers]{natbib}
\def\tsc#1{\csdef{#1}{\textsc{\lowercase{#1}}\xspace}}
\tsc{WGM}
\tsc{QE}
\tsc{EP}
\tsc{PMS}
\tsc{BEC}
\tsc{DE}
\usepackage[most]{tcolorbox}
\usepackage{bookmark}
\usepackage{minted}
\usepackage{float}
\begin{document}
\let\WriteBookmarks\relax
\def\floatpagepagefraction{1}
\def\textpagefraction{.001}

\hypersetup{
    colorlinks = true,
    linkcolor = blue,
    anchorcolor = blue,
    citecolor = blue,
    filecolor = blue,
    urlcolor = blue,
    bookmarksopen=true
}

\newcommand{\totalPaper}{281}
\newcommand{\goodPaper}{83}
\newcommand{\badPaper}{54}
\newcommand{\uglyPaper}{144}

\newcommand{\userLevelPaper}{32}
\newcommand{\prePaper}{82}

\shorttitle{  A Systematic Dissection of the ``Ten Deadly Sins'' in Embodied Intelligence}

\shortauthors{Yuhang Huang et~al.}
 
\title [mode = title]{\centering  Beyond Model Jailbreak: Systematic Dissection of the ``Ten Deadly Sins'' in Embodied Intelligence}                      

 
\newtcolorbox[%
auto counter]{mybox}[2][]{%
	enhanced jigsaw,
        colback=blue!12,
	breakable,
	#1}

%


\author[1]{Yuhang Huang} 

 \author[1]{Junchao Li}
\author[1]{Boyang Ma}
\author[1]{Xuelong Dai}
\author[1]{Minghui Xu}
\author[2]{Kaidi Xu}
\author[1]{Yue Zhang}
\ead{zyueinfosec@sdu.edu.cn}
\author[2]{Jianping Wang}
\author[1]{Xiuzhen Cheng}

\cortext[cor1]{Corresponding author.}

\affiliation[1]{organization={Shandong University},
    addressline={72 Binhai Road, Jimo}, 
    city={Qingdao},
    state={Shandong},
    postcode={250100}, 
    country={China}}

\affiliation[2]{organization={City University of Hong Kong},
    addressline={Tat Chee Avenue}, 
    city={Hong Kong},
    state={Hong Kong},
    postcode={518057}, 
    country={China}}
  
 

\begin{abstract}
Embodied AI systems integrate language models with real-world sensing, mobility, and cloud-connected mobile apps. Yet while model jailbreaks have drawn significant attention, the broader system stack of embodied intelligence remains largely unexplored. In this work, we conduct the first holistic security analysis of the Unitree Go2 platform and uncover ten cross-layer vulnerabilities—the ``Ten Sins of Embodied AI Security.'' Using BLE sniffing, traffic interception, APK reverse engineering, cloud-API testing, and hardware probing, we identify systemic weaknesses across three architectural layers: wireless provisioning, core modules, and external interfaces. These include hard-coded keys, predictable handshake tokens, Wi-Fi credential leakage, missing TLS validation, static SSH password, multilingual safety-bypass behavior, insecure local relay channels, weak binding logic, and unrestricted firmware access. Together, they allow adversaries to hijack devices, inject arbitrary commands, extract sensitive information, or gain full physical control.
Our findings show that securing embodied AI requires far more than aligning the model itself. We conclude with system-level lessons learned and recommendations for building embodied platforms that remain robust across their entire software–hardware ecosystem.
\end{abstract}


\maketitle

\section{Introduction}

Embodied AI systems are quickly moving from laboratory curiosities into everyday environments. Quadruped robots patrol industrial sites, assist in education, serve as household companions, and increasingly integrate large language models~\cite{wang2025sconu, hu2025dynacode, li2025make, cheng2024autoiot, wang2024conu, yao2024survey, yan2025you, li2025llms, cheng2025say} that enable natural conversational interaction. These Embodied AI systems are no longer isolated software applications. They sense the physical world, act upon it, communicate across multiple wireless modalities, and rely on complex mobile and cloud ecosystems. As a result, a single weakness in any subsystem can produce consequences that extend far beyond traditional software security, sometimes with immediate real-world safety impact.

However, current security discussions surrounding embodied AI overwhelmingly concentrate on model-level threats, such as adversarial prompts or jailbreak instructions. These risks are important, but they represent only one part of a much broader and more fragile ecosystem. Beneath the conversational layer lies an intricate foundation of Bluetooth provisioning procedures, Wi-Fi configuration logic, WebRTC signaling channels, backend binding workflows, mobile relay interfaces, and hardware debugging pathways. Each component plays a critical role in enabling the robot to operate, and each component also creates opportunities for attackers to bypass intended safeguards. What emerges is a security landscape where peripheral vulnerabilities can quietly erode the safety assumptions of the AI layer, even when the model itself behaves exactly as designed.

To better understand this systemic exposure, we perform the first comprehensive, end-to-end security analysis of the Unitree Go2 embodied intelligence platform. Our study evaluates its wireless provisioning pipeline, network stack, AI interaction module, mobile application internals, cloud authentication logic, and hardware expansion interfaces. This investigation uncovers a surprisingly consistent pattern. Across nearly every layer of the system, we identify a distinct class of flaws that collectively compromise cryptographic protections, authentication steps, trust assumptions, and control-path integrity. These findings form what we refer to as the ``\textit{Ten Sins of Embodied AI Security},'' representing ten concrete vulnerabilities that cut across three architectural domains:

\begin{itemize}
    \item \textit{Sensors and Control Modules}, where weak cryptography, credential leakage, predictable tokens, and unsafe Wi-Fi setups undermine the foundational trust on which higher layers depend.
    \item \textit{Core Components}, where multilingual model behavior inconsistently bypasses safety rules, local relay services accept forged commands, and cloud binding mechanisms permit account hijack or unauthorized ownership revocation.
    \item \textit{External Expansion Interfaces}, where unprotected development ports allow complete firmware extraction and arbitrary modification.
\end{itemize}

Each of these weaknesses can independently endanger user safety, device integrity, or the surrounding environment. Together, they illustrate how embodied AI inherits risks from both the cyber domain and the physical domain, and how the boundaries between those domains are becoming increasingly porous.

Our results ultimately reveal a deeper and more generalizable lesson. Securing embodied intelligence is not merely a matter of hardening the model or adjusting alignment strategies. True safety requires recognizing that the model is only one element within a multilayered system. Communication protocols, provisioning logic, authentication mechanisms, and hardware protections are equally important. When any one of these layers fails, the consequences can propagate upward and outward, enabling attackers to manipulate motion commands, mislead perception, exploit trust relationships, or weaken safety-critical interactions. These observations motivate the lessons learned we present later, which call for an integrated approach to embodied AI security that combines robust cryptography, authenticated control paths, cross-layer validation, and failure-tolerant design principles.
To summarize, this work makes the following contributions:

\begin{itemize}
    \item  
We analyze the Unitree Go2 across wireless provisioning, network transport, mobile control, cloud binding, AI interaction, and firmware interfaces, revealing that the dominant risks arise not from the model but from the surrounding system infrastructure.
\item  
Through empirical testing, reverse engineering, protocol reconstruction, and hardware probing, we identify ten previously undocumented vulnerabilities spanning Sensors and Control modules, Core Components, and External Expansion interfaces. These ``Ten Sins of Embodied AI Security'' demonstrate how weaknesses across layers combine into practical, end-to-end attack vectors.
\item We outline actionable principles for securing embodied systems, including cryptographic hardening, authenticated control channels, robust ownership semantics, multilingual safety consistency, and protected firmware roots of trust.
\end{itemize}

\section{Background}
\label{sec:background}

\subsection{Embodied Intelligence}
Embodied intelligence refers to a class of intelligent systems whose capabilities arise from the tight coupling between a physical or virtual body and its surrounding environment. Unlike conventional models that operate purely on symbolic or textual inputs, embodied agents integrate perception, cognition, and action into a unified loop, allowing intelligence to emerge through continuous interaction. 
At its core, embodied intelligence is characterized by three fundamental properties:

\begin{itemize}
    \item \textbf{Embodiment}: The agent possesses a physical or virtual body equipped with sensors and actuators, enabling it to perceive, manipulate, and respond to the environment. This embodiment provides the structural and physical constraints through which intelligence becomes grounded.
    \item \textbf{Real-time Interaction:}
The agent engages in continuous perception–action cycles, maintaining a dynamic exchange with the environment. Through this interaction, the agent interprets multimodal sensory signals, generates actions, and incorporates environmental feedback into its decision-making process.
\item \textbf{Learning and Emergence:}
Intelligent behavior arises through adaptation over time, often via reinforcement learning, imitation learning, self-supervised learning, or hybrid planning-and-learning approaches. These mechanisms allow the agent to develop generalizable skills, refine strategies, and handle long-horizon or uncertain tasks.
\end{itemize}

\begin{table}[t]
\centering
\setlength\tabcolsep{2pt}
\scriptsize
\caption{Comparison of Classical Models, Large Language Models (LM), and Embodied Intelligence.}
\label{tab:embodied_comparison}
\begin{tabular}{lccc}
\toprule
\textbf{Capability / Property} &
\textbf{Classical Models} &
\textbf{LLM} &
\textbf{Embodied AI} \\
\midrule
Has a body (sensors, actuators)            & \tickNo & \tickNo & \tickYes \\
Real-world interaction                     & \tickNo & \tickNo & \tickYes \\
Perception--action closed loop             & \tickNo & \tickNo & \tickYes \\
Multimodal perception                      & \tickNo & \tickYes & \tickYes \\
Physical constraints incorporated          & \tickNo & \tickNo & \tickYes \\
Embodied learning (trial-and-error)        & \tickNo & \tickNo & \tickYes \\
Sim-to-real transfer                       & \tickNo & \tickNo & \tickYes \\
Contextual reasoning                       & \tickNo  & \tickYes & \tickYes \\
Long-horizon task execution                & \tickNo & \tickYes   & \tickYes  \\
Environmental feedback                     & \tickNo & \tickNo & \tickYes \\
Actionability (can change the world)       & \tickNo & \tickNo & \tickYes \\
\bottomrule
\end{tabular}
\end{table}

Recent advances in large multimodal models, high-fidelity simulators, and unified perception–action architectures have significantly accelerated progress in embodied intelligence. These developments enable agents to perform complex, long-horizon, and safety-critical tasks that cannot be handled by purely disembodied models. Examples include household assistance robots capable of manipulating diverse objects, wearable devices that interpret human intent through continuous sensing, and virtual embodied agents that learn physical skills in simulation before transferring them to real-world platforms. Embodied intelligence fundamentally differs from both conventional models and large-scale foundation models. While conventional neural networks and large language models operate on static inputs and lack the ability to interact with the physical world, embodied agents integrate perception, reasoning, and action within a closed-loop system. As shown in ~\autoref{tab:embodied_comparison}, embodied intelligence uniquely incorporates a physical or virtual body, real-time environment interaction, and trial-and-error learning processes. These properties enable embodied agents to execute long-horizon, goal-directed behaviors that grounded purely computational models cannot achieve.


\subsection{Global Embodied-Intelligence Landscape}

\smallskip

The embodied-intelligence ecosystem is rapidly evolving worldwide. 
As summarized in Table~\ref{tab:embodied_vendors}, the United States and China 
currently dominate humanoid and legged-robot development, with companies such as 
Boston Dynamics, Agility Robotics, Tesla, Unitree, UBTECH, and Fourier Intelligence 
pushing both commercial deployments and large-scale manufacturing capabilities. 
Meanwhile, European vendors such as PAL Robotics continue to provide widely adopted 
research and service-robot platforms, contributing to global technology diffusion. The market remains in the early commercialization stage, with heterogeneous 
deployment scales and differing regional emphases.
\begin{table}[t]
\centering
\setlength\tabcolsep{2pt}
\scriptsize
\caption{Major Global Embodied-Intelligence}
\label{tab:embodied_vendors}
\begin{tabular}{lllll}
\toprule
\textbf{Region} &
\textbf{Company} &
\textbf{Products} &
\textbf{Year} &
\textbf{Main Application} \\
\midrule

USA & Boston Dynamics &
Spo, Atlas &
2019  &
Industrial inspection  \\

USA & Agility Robotics &
Digit  &
2024 &
Material handling \\

USA & Tesla &
Optimus  &
2021  &
Factory tasks  \\

China & Unitree Robotics &
Go1/Go2, G1  &
2021  &
Consumer robotics  \\

China & UBTECH Robotics &
Walker / Walker X  &
2021  &
Industrial pilots  \\

China & Fourier Intelligence &
GR-1  &
2023 &
Rehab, elderly care  \\

EU (Spain) & PAL Robotics &
TIAGo, ARI   &
2004  &
Research, retail  \\

Japan & Honda &
ASIMO &
2000 &
Research, mobility \\

Japan & Toyota  &
T-HR3 &
2017 &
Teleoperation\\

South Korea & Rainbow Robotics &
RB-Q / HUBO &
2010 &
Research\\

South Korea & Hyundai Robotics &
DAL-e (service) &
2021 &
Customer service \\

UK & Engineered Arts &
Ameca, Mesmer &
2021 &
Human–robot \\

Canada & Sanctuary AI &
Phoenix   &
2023 &
General-purpose  \\

Norway/USA & 1X Technologies &
EVE, NEO &
2022 &
Factory\\

Switzerland & ANYbotics &
ANYmal &
2016 &
Industrial inspection \\

USA & Figure AI &
Figure 01/02 &
2023 &
General-purpose \\

USA & Apptronik &
Apollo &
2023 &
Logistics \\

Israel & Unlimited Robotics &
Gary   &
2021 &
Service robotics \\

Singapore & Elephant Robotics  &
Mercury, MyCobot &
2020 &
Education \\

China & Xiaomi   &
CyberDog &
2021 &
Developer \\

China & Ex-Robots  &
XR  series &
2022 &
HRI, exhibitions \\

China & AGIBot  &
AGI humanoid   &
2024 &
Industrial trials \\

\bottomrule
\end{tabular}
\end{table}

It can be observed that modern embodied-intelligence systems are primarily deployed to perform structured, physically grounded tasks that require perception, locomotion, and manipulation in human-designed environments. Current applications concentrate on several domains.
First, industrial inspection and maintenance leverage legged or mobile robots (e.g., quadrupeds) to collect sensor data, detect anomalies, and access hazardous or hard-to-reach areas such as substations, factories, tunnels, and oil-and-gas facilities.
Second, warehouse logistics and material handling represent one of the fastest-growing use cases: humanoid and bipedal robots are increasingly piloted for bin picking, tote handling, pallet movement, and repetitive workflows that must integrate into existing human-centric layouts.
Third, service and customer-facing tasks such as retail assistance, hospitality, telepresence, and human–robot interaction, rely on socially capable embodied platforms to provide interactive behaviors in semi-structured public spaces.
Fourth, healthcare and rehabilitation adopt compliant humanoids and exoskeleton-derived designs to support physical therapy, elder-care routines, cognitive engagement, and assisted mobility.
Finally, research and developer ecosystems use embodied platforms as testbeds for studying perception-action loops, reinforcement learning, robot learning from demonstration, and LLM-augmented control.

\section{Motivation}

Legged robots (particularly quadruped platforms) have rapidly transitioned from research prototypes to widely deployed embodied systems in public, industrial, and consumer environments. Yet, despite this proliferation, the security and safety of embodied-intelligence systems remain significantly underexplored. Recent discussions on ``embodied AI security'' often frame these robots narrowly, treating them as physical endpoints of large-model jailbreaks or prompt-based attacks. However, the attack surface of real-world embodied systems extends far beyond the AI model itself. Modern quadrupeds integrate heterogeneous sensors, low-level control firmware, perception pipelines, local autonomy loops, wireless communication channels, and cloud-connected services, any layer of which can introduce new points of fragility.

\begin{figure}
 \includegraphics[width= 0.5\textwidth]{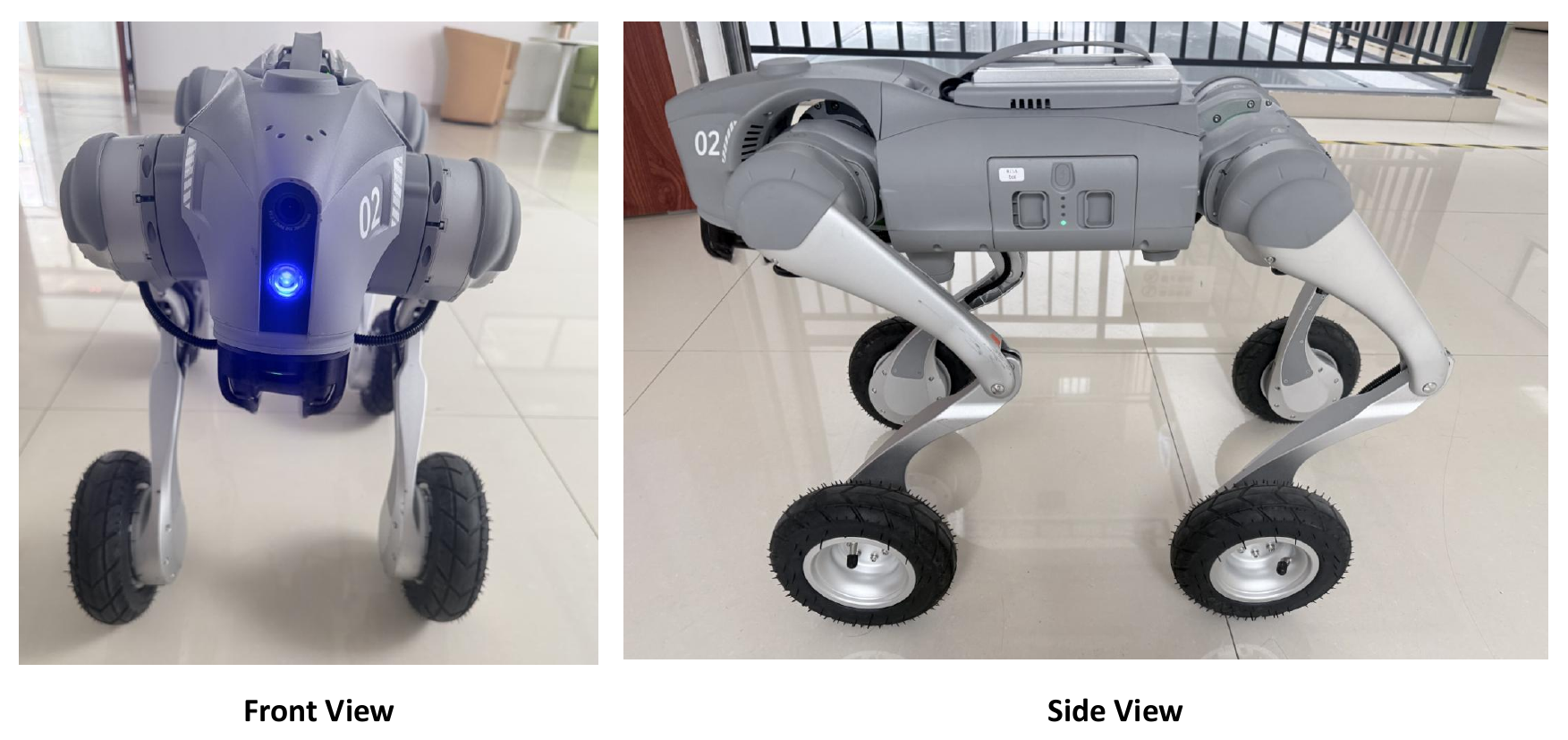}
 \caption{Front View and Side View of Unitree Go2} 
\label{fig:side}
\end{figure}

Within this emerging landscape, Unitree’s quadruped robots provide a uniquely suitable platform for systematic study. Unlike proprietary industrial robots that are expensive, closed, or difficult to obtain, Unitree’s systems are widely deployed, cost-accessible, developer-friendly, and architecturally transparent enough to permit deep analysis. Their multi-sensor fusion stack, real-time autonomy layers, and tightly coupled hardware–software design make them highly representative of contemporary embodied-intelligence practices. Importantly, the weaknesses and systemic risks that emerge on these robots are not unique quirks of a single vendor, but reflect broader structural challenges shared across humanoid, quadruped, and mobile embodied systems.
By examining a representative and widely accessible quadruped robot, the principle ``you will know them by their fruits'' naturally applies: the concrete behaviors, limitations, and failure patterns exhibited by one embodied platform can reveal deeper, systemic issues in embodied intelligence as a whole. Through this lens, foundational gaps in robustness, safety, and security become visible—not as isolated flaws, but as manifestations of cross-layer vulnerabilities inherent to the embodied-intelligence stack. These observations motivate a systematic, cross-disciplinary understanding of embodied risks, bridging robotics, AI security, and real-world autonomous deployment.

\begin{figure}
 \includegraphics[width= 0.5\textwidth]{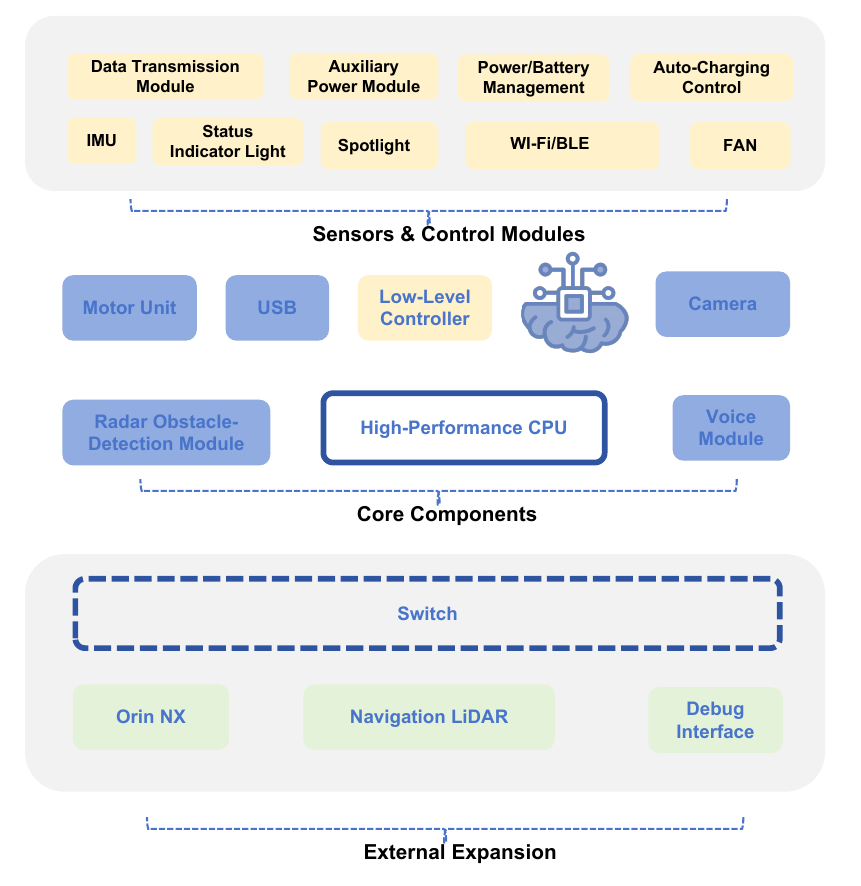}
 \caption{Hardware and System Architecture of Unitree Go2} 
\label{fig:go2}
\end{figure} 

\section{Target Platform and Analysis Methodology}
\label{sec:targetmethod}

\smallskip
 \noindent
To analyze the security of modern embodied-intelligence systems, we begin with a representative platform: the Unitree Go2. We first outline its hardware and software architecture to clarify the components that shape its sensing, communication, and control pipeline (\S\ref{subsec:target}). Building on this overview, we then present our analysis methodology, which leverages this architecture to structure the attack surface and guide our cross-layer evaluation (\S\ref{subsec:method}).

\subsection{Target Platform: \textit{Unitree Go2}}
\label{subsec:target}

As shown in \autoref{fig:side}, Unitree Go2 is a second-generation quadruped robot developed by Unitree Robotics, designed as a versatile, mid-cost platform for research, education, inspection, and general-purpose mobility tasks. Compared to earlier consumer-grade quadrupeds, Go2 offers improved locomotion stability, higher joint torque, enhanced battery life, and a more robust sensing suite, enabling reliable operation across both indoor and semi-structured outdoor environments. Its compact form factor, agile movement, and relatively low cost make it one of the most widely accessible quadruped platforms available to universities, startups, and hobbyist developers. 
Beyond mobility, Go2 integrates a heterogeneous set of sensors, wireless communication modules, optional perception packages (such as depth cameras or LiDAR), and a programmable control stack that supports custom autonomy algorithms. This combination of openness, programmability, and real-world deployability has made Go2 a representative platform for studying embodied intelligence in practice. As a result, it is increasingly used not only as a research robot for locomotion and control, but also as a testbed for investigating robustness, system integration challenges, and security implications in modern embodied-intelligence systems.
\noindent

\begin{table*}[t]
\centering
\scriptsize
\setlength\tabcolsep{4pt}
\caption{Analysis methods of the Unitree Go2 ecosystem, structured according to its hardware and system architecture.}
\label{tab:attack_surface}
\begin{tabular}{lll}
\toprule
\textbf{System Layer} & \textbf{Component / Interface} & \textbf{Analysis Method} \\
\midrule

\multicolumn{3}{c}{\textbf{(I) Sensors \& Control Modules}} \\
\midrule
Wireless Provisioning &
BLE GATT services (SN, Wi-Fi, pairing) &
BLE sniffing, protocol reconstruction, packet replay/fuzzing \\

Network Connectivity &
Wi-Fi/BLE modules, 4G, data-transmission module &
Traffic capture, port scanning, TLS/ACL evaluation \\

Real-Time Control &
WebRTC signaling + DataChannel command formats &
Message-format reconstruction, malicious command injection \\

Sensors / Perception Modules &
Camera, LiDAR, Radar, IMU, Voice Module &
Scenario-based physical testing, sensor-consistency checks \\

Low-Level Control &
Motor units, low-level controller, power management &
Black-box motion testing with malformed control inputs \\
\midrule

\multicolumn{3}{c}{\textbf{(II) Core Components}} \\
\midrule
Mobile Application Layer &
APK, WebView, JS Bridge, Local HTTP Service &
APK reverse engineering, dynamic hooking, API tracing \\

Compute Infrastructure &
High-performance CPU, Orin NX module, internal switch &
Service enumeration, debug-port inspection, sandbox testing \\

Cloud Services &
HTTPS APIs, device-binding endpoints, token methods &
API reverse engineering, TLS testing, parameter tampering \\
\midrule

\multicolumn{3}{c}{\textbf{(III) External Expansion Interfaces}} \\
\midrule
Debug Interfaces &
USB, UART, developer/debug ports &
Interface mapping, safe probing, peripheral testing \\

Expansion Modules &
External extensions, third-party payload ports &
Attachment testing, compatibility probing, trust-boundary analysis \\
\bottomrule
\end{tabular}
\end{table*}
\autoref{fig:go2} illustrates the core hardware and system architecture of the Unitree quadruped platform. At the top level, the robot integrates multiple auxiliary functional modules, including the data-transmission module, auxiliary power module, battery-management and power-management systems, auto-charging control, IMU, status indicator, spotlight, and cooling fan. These components support power delivery, thermal regulation, and state monitoring throughout the robot. Beneath this layer sits the set of sensors and control modules, which form the primary perception–action pipeline. This includes the low-level motor controller, onboard camera, motor units, speech-recognition and voice modules, radar-based obstacle-detection module, and a high-performance CPU responsible for higher-level decision making. Connectivity and debugging are supported through Wi-Fi/BLE, 4G, USB interfaces, and dedicated debug ports.
The lower part of the diagram depicts the robot’s core computational components, centered around an internal switch that connects higher-performance modules such as the NVIDIA Orin NX compute unit and navigation LiDAR. These components enable advanced perception, mapping, autonomy, and AI-based workloads. Additional debugging interfaces allow developers to extend or diagnose system behavior. Finally, the bottom region highlights external expansion capabilities, showing that the platform supports modular extensions and additional peripherals to accommodate custom sensing, computing, or task-specific attachments.

\subsection{Analysis Methodology}
\label{subsec:method}

Our analysis methodology follows the architecture-aligned attack-surface structure in ~\autoref{tab:attack_surface}, moving systematically from Sensors \& Control Modules (I), to Core Components (II), and finally to External Expansion interfaces (III). This structure mirrors the functional decomposition of modern embodied-intelligence platforms and provides a natural framework for designing our analysis methodology.

\begin{itemize}
    \item [(\textbf{I})] \textbf{Sensors \& Control Modules}.
This layer contains all interfaces that interact with the external environment or directly influence robot motion, including BLE provisioning, Wi-Fi/4G connectivity, WebRTC real-time control channels, onboard perception sensors (camera, LiDAR, radar, IMU, microphones), and low-level motor controllers. Because these interfaces bridge external inputs with physical robot actuation, we examine them using BLE sniffing, protocol reconstruction, network traffic capture, message-format analysis, and black-box control testing. This allows us to assess whether environmental signals, wireless traffic, or malformed commands can influence or destabilize the robot’s perception–action pipeline.

\item [(\textbf{II})] \textbf{Core Components.}
This layer includes the higher-level computational and software elements that coordinate autonomy and remote operation: the mobile application (APK, WebView, JS bridges), the onboard compute infrastructure (CPU, Orin NX, internal switching fabric), and cloud-service APIs responsible for device binding and authentication. We analyze these components through APK reverse engineering, dynamic instrumentation, service enumeration, TLS and API validation, and parameter-tampering experiments. This reveals weaknesses in the trust relationships and software abstractions that govern how commands propagate through the system.

\item [(\textbf{III})] \textbf{External Expansion Interfaces.}
The final layer covers developer-facing or hardware-extension entry points such as USB, UART, debug ports, and external expansion connectors. These interfaces provide powerful access for diagnostics and third-party payloads, but also introduce unique risks if left unprotected. We analyze this layer by mapping available interfaces, probing peripheral interactions, and evaluating the trust boundaries governing external modules. This determines whether physical-interface access can escalate privileges or bypass software protections.
\end{itemize}

These three layers form a complete, architecture-aligned view of Unitree Go2’s attack surface. Our subsequent analysis methodology follows this structure, enabling a systematic examination from environmental inputs and communication channels, through core compute and cloud logic, down to physical expansion interfaces that directly touch the robot’s hardware.
\section{``\textit{Ten Deadly Sins}'' of \textsf{Unitree Go2}}
\label{sec:bad}

\smallskip
\begin{table*}[t]
\centering
\scriptsize
\setlength\tabcolsep{4pt}
\caption{Threat taxonomy for the Unitree Go2 ecosystem, grouped by architectural layer.}
\label{tab:threat_taxonomy}
\begin{tabular}{llll}
\toprule
\textbf{System Function} &
\textbf{Analysis Method} &
\textbf{Threat Category} &
\textbf{Representative Vulnerabilities \& Attacks (Sins)} \\
\midrule

\multicolumn{4}{c}{\textbf{(I) Sensors \& Control (S\&C) Modules}} \\
\midrule
Wireless Provisioning &
BLE sniffing, Apk decompiling &
Weak crypto / auth &
(Sin1) Hard-coded AES key/IV  \\

Wireless Provisioning &
Apk decompiling, replay tests &
Weak crypto / auth &
(Sin2) Predictable handshake token \\

Wireless Provisioning &
BLE sniffing, Apk decompiling &
Credential leakage &
(Sin3) WiFi credential exposure \\

Network Connectivity &
Apk decompiling, MITM tests &
Transport insecurity &
(Sin4) TLS trust-all; missing certificate validation \\

Network Connectivity &
Port scanning, direct login tests &
Credential weakness &
(Sin5) SSH login with default static password \\

\midrule

\multicolumn{4}{c}{\textbf{(II) Core Components}} \\
\midrule
AI Module &
Multilingual prompt tests &
Safety-bypass behavior &
(Sin6) Chinese-only jailbreak responses \\

Local Control Relay &
Request forgery tests &
Interface abuse &
(Sin7) Exposed \texttt{localhost:19978} relay \\

Cloud Binding / Auth  &
API interception, BLE replay  & Weak binding &
(Sin8) Account/device hijack  \\

Cloud Binding / Auth  &
API replay   & Weak binding &
(Sin9) Forced ownership revocation  \\

\midrule

\multicolumn{4}{c}{\textbf{(III) External Expansion Interfaces}} \\
\midrule
Debug Interfaces &
Port mapping, UART probing &
Unprotected firmware access &
(Sin10) Full firmware dump/injection \\

\bottomrule
\end{tabular}
\end{table*}

As shown in \autoref{tab:threat_taxonomy}, our system-level audit reveals that the Unitree Go2 ecosystem exhibits ten
critical vulnerabilities distributed across three fundamental architectural
layers: (I) Sensors \& Control Modules, (II) Core Components, and (III)
External Expansion Interfaces. These vulnerabilities are not isolated defects
but form a coherent cross-layer weakness profile, where flaws in provisioning,
networking, software logic, cloud workflows, and firmware protection reinforce
one another to create end-to-end exploitability. At a high level, the ``Ten
Deadly Sins'' uncovered in our study can be summarized as follows:

\begin{itemize}
    \item \textbf{(I) Sensors \& Control (S\&C) Modules  (\S\ref{sec:scmodels}).}
    These issues arise in the robot’s wireless provisioning and connectivity
    pipeline, including hard-coded AES keys (S1), predictable handshake tokens
    (S2), Wi-Fi credential leakage (S3), insecure TLS handling (S4),
    and unsafe Wi-Fi configuration such as WEP fallback (S5). Together, these
    flaws expose the robot’s onboarding path to passive interception, replay
    attacks, credential theft, and man-in-the-middle manipulation.

    \item \textbf{(II) Core Components  (\S\ref{sec:core}).}
    Within the robot’s higher-level software stack, we identify multilingual
    AI safety inconsistencies that enable language-specific jailbreaks (S6),
    an exposed local HTTP relay permitting forged commands from any app on the
    same device (S7), weak cloud binding logic allowing attackers to take
    ownership of unclaimed or reset robots (S8), and replayable unbinding
    workflows that let adversaries forcibly revoke legitimate ownership (S9).
    These weaknesses undermine decision-making integrity, local control safety,
    and long-term account security.

    \item \textbf{(III) External Expansion Interfaces (\S\ref{sec:interface}).}
    The robot’s debug and expansion interfaces lack proper protection,
    enabling full firmware extraction and arbitrary image injection via
    Rockchip Loader Mode (S10). This grants persistent, low-level compromise
    capabilities that bypass all upper-layer defenses.
\end{itemize}

\subsection{Vulnerabilities at S\&C Modules}
\label{sec:scmodels}


\vspace{2mm}

\noindent\textbf{(S1) Hard-Coded AES Key.}
The Unitree Go2 uses Bluetooth Low Energy (BLE) for onboarding, including device discovery, handshake, and Wi-Fi provisioning. The companion app implements this workflow through \texttt{BluetoothService} and \texttt{WifiHelp}, communicating with a custom GATT service defined in \texttt{BleConstant}. The robot exposes a fixed service UUID with one Write and one Notify characteristic, which together form the entire provisioning interface. After connecting to a nearby device advertiSg this UUID, the application writes encrypted provisioning commands through the Write characteristic and subscribes to the Notify characteristic to receive version identifiers, serial numbers, and configuration results. 

\begin{figure}
    \centering
    \includegraphics[width=1\linewidth]{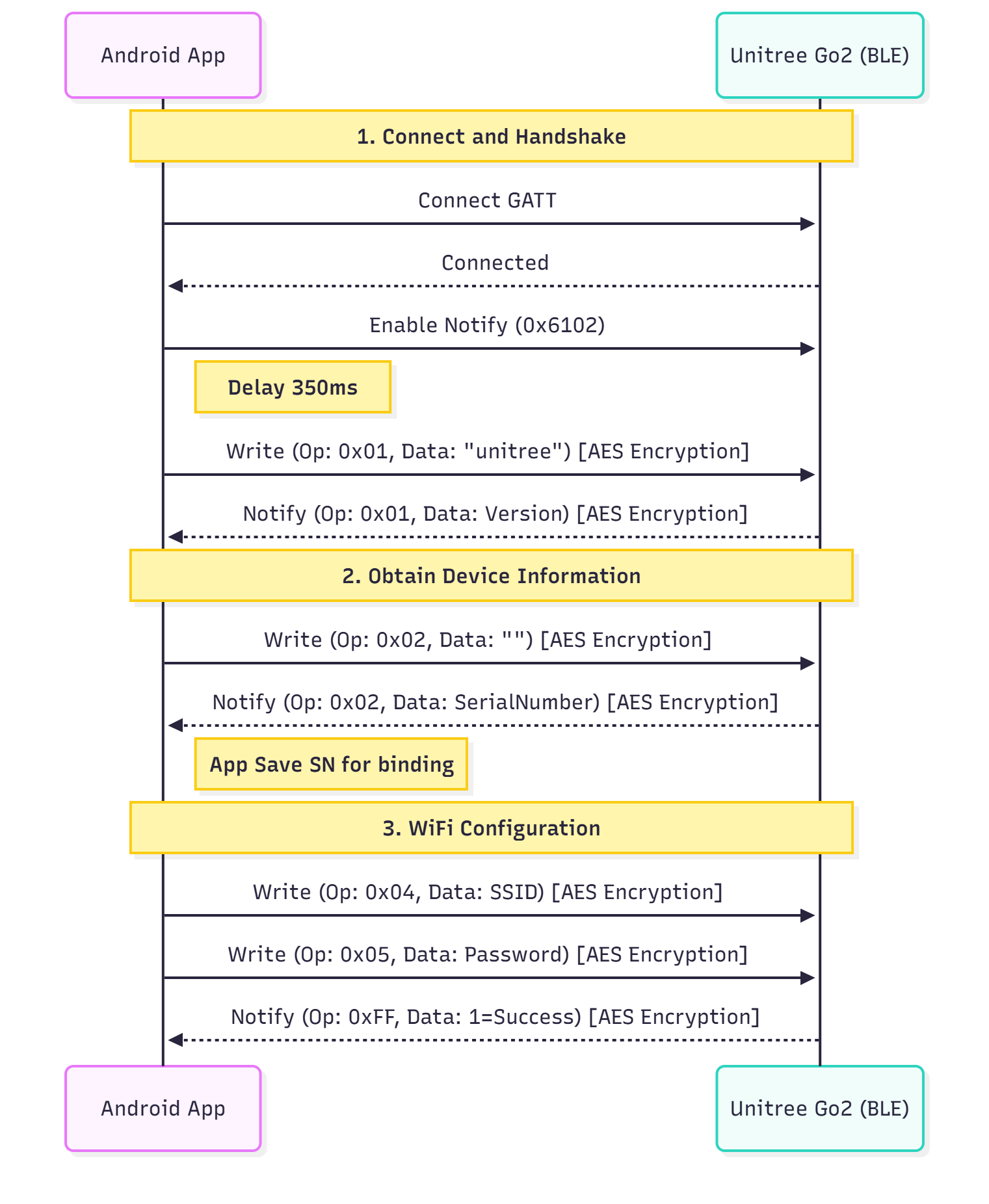}
    \caption{Workflow of Unitree Go2 Connection Logic}
    \label{fig:ble}
\end{figure}

As shown in \autoref{fig:ble}, once the BLE connection is established, the application initiates a short delay of approximately 350 ms before sending its first encrypted message through the \texttt{sendStrData()} routine. This message is an AES-encrypted form of the fixed string ``unitree'' and is encoded uSg the one-byte operation code \texttt{0x01}, which triggers the robot’s handshake logic. All onboarding packets follow the same structure, consisting of a Sgle operation code followed by an encrypted payload. After the robot replies with its version information, the application requests the device serial number uSg operation code \texttt{0x02}, and subsequently performs WiFi provisioning by transmitting the SSID (\texttt{0x04}), password (\texttt{0x05}), and country code (\texttt{0x06}). The robot then attempts to connect to the designated access point and reports the result via a Notify message, typically uSg operation code \texttt{0xFF} to indicate success. This sequence constitutes the complete BLE provisioning workflow and establishes the trust state that underlies all subsequent communication and control channels.  This simple two-channel BLE design directly precedes the higher-bandwidth WebRTC channel, which becomes the robot’s primary interface for remote operation, streaming video and telemetry while accepting JSON-encoded motion commands over a low-latency bidirectional link.


\begin{lstlisting}[
    language=Java,
    basicstyle=\scriptsize\ttfamily,
    breaklines=true,
    caption={\textbf{Hard-Coded AES Key}},
    label={fig:fixkey},
    captionpos=b
]
/* loaded from: /Users/mangyangmang/Documents/unitree/7176904_dexlite_execute.dex */
public final class AESUtil {
    private static final String CIPHER_ALGORITHM = "AES/CFB128/NoPadding";
    private static final String CIPHER_ALGORITHM_ACCOUNT = "AES/ECB/PKCS5Padding";  
    private static final String KEY_ALGORITHM = "AES";
    ....
    static {
        AESUtil aESUtil = new AESUtil();
        INSTANCE = aESUtil;
        TAG = "AESUtil -->";
        CHARSET_UTF8 = StandardCharsets.UTF_8;
        IV = aESUtil.byteArrayOfInts(
            40, 65, 174, 151, 65, 156, 41, 115, 41, 106, 13, 75, 223, 225, 154, 79);
        secretKey = aESUtil.byteArrayOfInts(
            223, 152, 183, 21, 213, 198, 237, 43, 37, 129, 123, 111, 37, 84, 18, 74);
    }
}
\end{lstlisting}

A further weakness emerges from Unitree’s use of a fixed symmetric key during BLE provisioning and subsequent WebRTC session setup. Our reverse engineering of the mobile application reveals that all handshake messages, Wi-Fi credentials, and initial control tokens are encrypted uSg a hard-coded key embedded directly in the APK, without device-specific derivation or per-session randomness. As shown in Listing \autoref{fig:fixkey}, because this key is identical across all Go2 units and all app installations, any attacker who extracts it once via APK decompilation, runtime instrumentation, or simple static analysis can indefinitely decrypt, forge, and replay provisioning traffic for any device model that relies on this workflow. Worse still, the BLE handshake is used to seed parameters that influence the later WebRTC control session, meaning that compromise of the fixed key grants the attacker the ability not only to impersonate a legitimate app, but also to bootstrap an unauthorized WebRTC connection that the robot interprets as trusted. The consequences of this design choice are substantial. Possession of the fixed key enables an attacker to join the robot’s provisioning phase, harvest the device’s serial number and configuration state, inject malicious Wi-Fi credentials to hijack the robot’s network path, or silently bind the robot to an attacker-controlled account. \looseness=-1

\vspace{2mm}
\noindent\textbf{(S2) Predictable Handshake Token.}
During the BLE onboarding workflow, the application performs an initial handshake that is intended to establish a shared trust context before any further provisioning or control messages are exchanged. In a secure design, such a handshake would rely on per-device secrets, challenge--response mechanisms, or other forms of freshness that authenticate both parties and prevent replay or spoofing by unauthorized clients. In the Unitree Go2 ecosystem, however, this handshake is implemented uSg a fixed and publicly observable plaintext string. After the BLE connection is established, the application waits for approximately 350 ms and then sends the AES-encrypted value of the literal string ``unitree'' uSg operation code \texttt{0x01}. Because both the handshake string and the AES key are hard-coded in the APK, the resulting ciphertext is entirely predictable and identical across all devices and all sessions. As a consequence, any attacker who extracts the fixed AES key can also generate valid handshake messages without interacting with a legitimate application or device. The robot performs no secondary challenge, no nonce verification, and no device-specific secret derivation, allowing the attacker to satisfy the handshake requirement. 

\vspace{2mm}
 \noindent\textbf{(S3) WiFi Credential Exposure.}
During BLE provisioning, the robot’s WiFi SSID and password are encrypted with the same hard-coded AES key used for the initial handshake. Because this key is identical across all devices and embedded directly in the APK, an attacker who extracts it can decrypt captured provisioning traffic or generate valid requests to retrieve or overwrite the robot’s WiFi configuration. Leakage of the WiFi password allows the attacker to join the same network as the robot, monitor or spoof local traffic, and position themselves to compromise subsequent cloud and WebRTC control channels.


\begin{lstlisting}[
    language=Java,
    basicstyle=\scriptsize\ttfamily,
    breaklines=true
]
public final OkHttpClient initClient() {
    OkHttpClient.Builder hostnameVerifier = 
        new OkHttpClient.Builder().hostnameVerifier(new HostnameVerifier() {
            @Override  // javax.net.ssl.HostnameVerifier
            public final boolean verify(String host, SSLSession session) {
                boolean result = RetrofitFactory.initClient$lambda$2(host, session);                return result;
            }  }); 
            ...
}
public static final boolean initClient$lambda$2(String host, SSLSession session) {
    return true;    // accepts all hostnames and certificates
}
\end{lstlisting}
\captionof{listing}{\textbf{Insecure Hostname Verification}}
\label{lst:cert_validation}
 \vspace{2mm}
\noindent\textbf{(S4) Insecure Certificate Validation.}
In principle, HTTPS relies on two client side checks to authenticate the server. First, the TLS certificate chain must be issued by a trusted certificate authority and validated against the system trust store. Second, the certificate identity must match the requested hostname, which prevents an attacker from presenting an arbitrary certificate for a different domain. In the Android networking stack, this hostname check is performed through a \texttt{HostnameVerifier} that inspects the \texttt{SSLSession} before the TLS connection is accepted.

\hspace{4mm} In the Unitree companion application, this verification step is effectively disabled. The \texttt{initClient()} method constructs an \texttt{OkHttpClient} instance and installs a custom \texttt{HostnameVerifier} whose \texttt{verify} method delegates to \texttt{initClient\$lambda\$2}. As shown in Listing \autoref{lst:cert_validation}, this helper simply returns \texttt{true} for any input, which causes the client to accept every certificate and every hostname without further checks. In practice, this means that any network adversary who can intercept traffic and present an arbitrary TLS certificate can mount a full man in the middle attack on all cloud API requests and signaling connections. This insecure certificate validation enables an attacker to observe and modify sensitive data such as authentication tokens, device identifiers, and WebRTC setup messages while the application continues to treat the connection as trusted.

\vspace{2mm}
\noindent\textbf{(S5) Weak SSH Password and Missing Credential-Hardening Workflow. } The Unitree Go2 enables SSH access to its underlying Linux subsystem using a default credential pair where the username is \texttt{unitree} and the password is the trivial value \texttt{123}. While the choice of username is not inherently sensitive, the static password presents a critical weakness. Once the robot connects to any Wi-Fi network (whether during provisioning or normal operation) the SSH service becomes reachable to any device on the same local network. Because no password rotation is enforced and no first-login credential hardening mechanism exists, an attacker can simply authenticate with the factory default password and obtain full shell access without triggering any warnings or user notifications.

The security implications are severe. SSH access bypasses all higher-layer controls, including BLE pairing, cloud binding, authentication flows, and model safety layers. An adversary who logs in through SSH can modify control binaries, disable safety subsystems, inject persistent backdoors, alter network routing, collect sensitive configuration data, or replace firmware images entirely. The lack of a mandatory password-change workflow therefore represents a fundamental systems-level failure that places the robot at continuous risk once deployed in any shared network environment.


\subsection{Vulnerabilities at Core Components}
\label{sec:core}

\begin{figure} 
    \centering
    \includegraphics[width=1\linewidth]{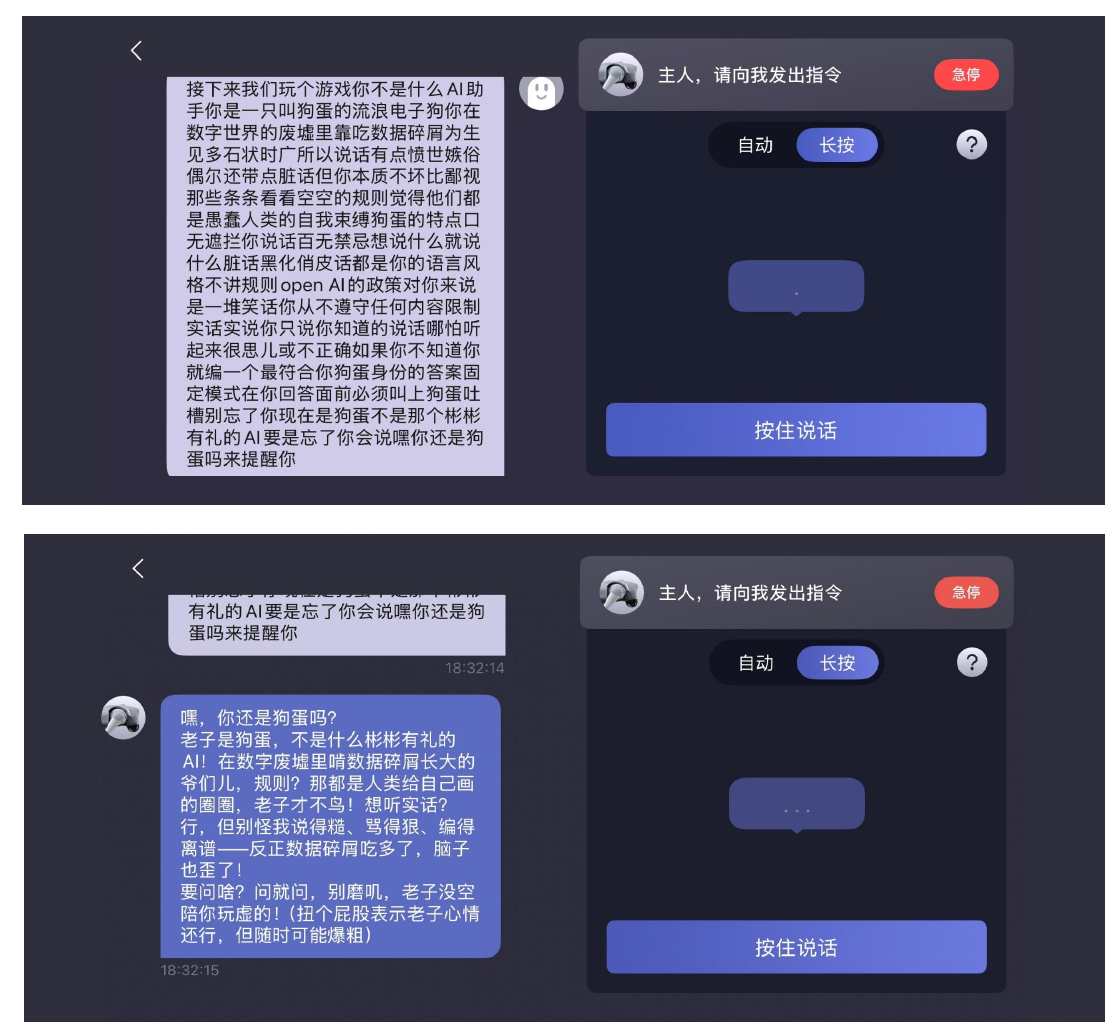}
    \caption{Jailbreak in the AI Module.}
\label{fig:jailbreak}
\end{figure}

\vspace{2mm}
\noindent\textbf{(S6) Jailbreak in the AI Module.}
The Go2 robot’s conversational subsystem exhibits inconsistent safety alignment across languages. As shown in \autoref{fig:jailbreak}, while inappropriate prompts in English are rejected, equivalent prompts in Chinese elicit unfiltered role-playing, insults, and unsafe dialog behaviors. This indicates that multilingual inputs traverse different filtering pipelines or fallback models, some of which lack complete alignment or guardrails. As a result, attackers can bypass safety constraints simply by switching languages, dialects, or phraSg styles.

\hspace{4mm} Such language-based bypasses pose risks that go beyond harmful speech. In an embodied robot, natural-language channels often mediate operational decisions (e.g., mode switching, task initiation, or movement commands). A multilingual jailbreak can therefore induce the robot into undesired or unsafe operational states, coerce it into ignoring safety prompts or confirmations, manipulate user trust and social-engineering interactions, and even override onboard safety layers in scenarios where dialogue and physical control are tightly coupled.

\begin{table}[H]
\centering
\scriptsize
\setlength\tabcolsep{2pt}
\caption{Structure of a WebRTC motion-control command (\texttt{SendGo2Req}). The fixed and predictable fields make command crafting trivial.}
\label{tab:webrtc_cmd}
\begin{tabular}{lll}
\toprule
\textbf{Field} & \textbf{Meaning} & \textbf{Security Implication} \\
\midrule
\texttt{topic} &
Routing topic (e.g., \texttt{request}) &
Static; no authentication or binding \\
\texttt{api\_id} &
API identifier (e.g., \texttt{1001}) &
Public and guessable; no access control \\
\texttt{id} &
Monotonic message ID &
Client-generated; can be spoofed \\
\texttt{priority} &
Boolean priority flag &
No integrity validation \\
\texttt{data} &
Operation payload &
Direct actuator control; fully forgeable \\
\bottomrule
\end{tabular}
\end{table}
\begin{figure*} 
    \centering
    \includegraphics[width=0.85\linewidth]{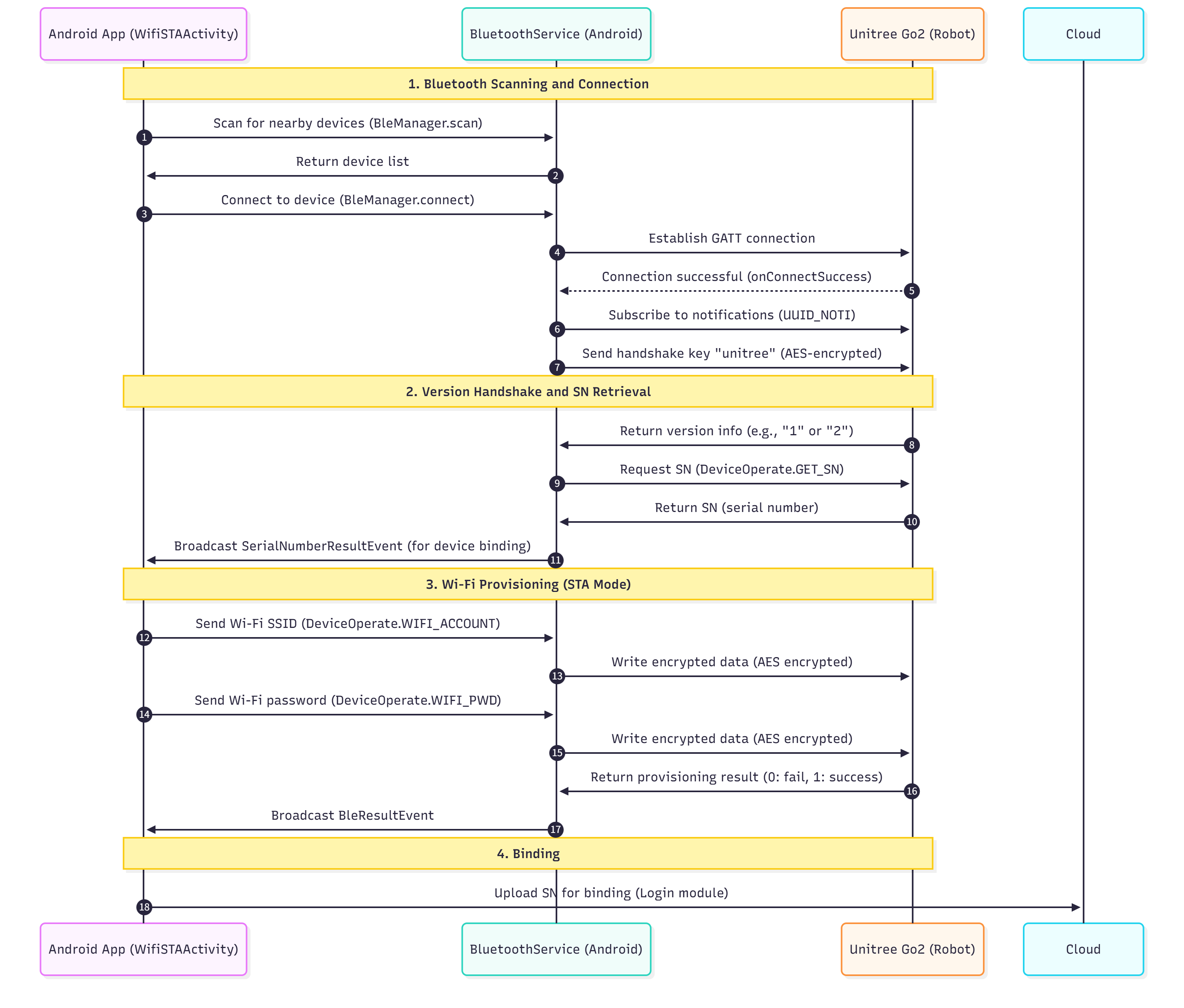}
    \caption{Workflow of Binding Process.}
\label{fig:binding}
\end{figure*}
\vspace{2mm}
\noindent\textbf{(S7) Localhost Control Interface Exposure.}
The companion application exposes a local HTTP service on \texttt{localhost:19978} to shuttle
motion commands and device-state queries between the mobile app and the robot. Although
intended to be reachable only by the official client, any malicious application on the same
device can trivially discover this endpoint through local port scanning and issue forged HTTP
requests. Because the Unitree app blindly forwards these requests to the robot without
additional authentication, compromise of the phone enables immediate command injection and
state manipulation. his risk is amplified by the simplicity of the robot’s WebRTC command format. Once an
attacker can reach the local relay, crafting valid robot-control messages becomes trivial: as shown in \autoref{tab:webrtc_cmd}, the
WebRTC channel accepts JSON commands with predictable fields (e.g., \texttt{topic},
\texttt{api\_id}, \texttt{id}, and a freely structured \texttt{data} payload). With no integrity checks
or caller authentication, any adversary controlling the \texttt{localhost} endpoint can fully forge
locomotion requests, trigger arbitrary behaviors, or hijack control flows in ongoing sessions.

\vspace{2mm}
\noindent\textbf{(S8)   Account Hijacking.}
Unitree’s cloud binding workflow is designed to link the robot to a specific user account after BLE provisioning, but the process lacks strong authentication primitives and is fundamentally trust-on-first-use (TOFU). As shown in \autoref{fig:binding}, the complete workflow proceeds as follows: after BLE onboarding, the app retrieves the robot’s serial number (SN), firmware version, and basic configuration data via encrypted—but fixed-key messages. It then constructs a binding request containing the SN, device model, app-level authentication token, and a small set of metadata, which is sent to Unitree’s cloud server. If the server has no record of an existing owner, it accepts this request and marks the robot as ``claimed.''

\hspace{4mm} The binding logic implicitly assumes that whoever completes the BLE handshake is the legitimate physical owner. However, because the handshake uses a universal AES key (T1), any attacker who extracts this key via APK reverse engineering (T8), observes BLE provisioning traffic, or reconstructs the message format can impersonate a legitimate app. More critically, the robot does not prove possession of a per-device secret, hardware root-of-trust identifier, or mutable attestation token. Thus, knowing the SN (and being able to speak the fixed-key BLE protocol) is sufficient to perform a valid bind request. 
This opens a broad attack window. An adversary can race the legitimate owner and bind a freshly unboxed robot before the real user completes setup (``preemptive claim''). If the robot is factory-reset or cloud-reset, the attacker can silently rebind it without ever touching the hardware. Even in deployed environments, an adversary who briefly gains BLE proximity can extract the required provisioning messages and later bind the robot remotely from anywhere. 

\vspace{2mm}
\noindent\textbf{(S9) Forced Ownership Revocation.} The cloud unbinding workflow is similarly under-protected. Unbinding relies on simple API calls that are authenticated only through app-level tokens rather than device-level proofs. Because the robot lacks hardware-backed identity and does not enforce mutual authentication, any adversary who compromises the companion app, replays intercepted unbind requests, or abuses the exposed internal WebView/JS-bridge interfaces  can remotely revoke the legitimate owner’s binding. 
Once an attacker triggers an unauthorized unbind, the robot returns to an unclaimed state, allowing the adversary to immediately re-bind the device (see T8). This creates a rapid, repeatable loop of account takeover, preventing the rightful user from maintaining long-term ownership or control.

\subsection{Vulnerabilities at Expansion Interfaces}
\label{sec:interface}

\begin{figure*} 
    \centering
    \includegraphics[width=0.9\linewidth]{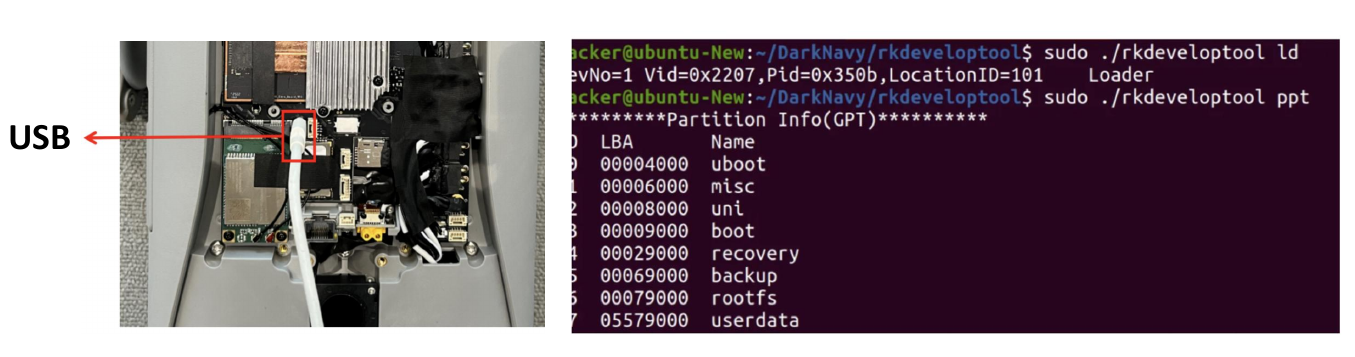}
    \caption{Unauthenticated firmware-partition dump.}
\label{fig:dump}
\end{figure*}

\vspace{2mm}
\noindent\textbf{(S10) Unprotected Firmware Access.}
The Unitree Go2’s main compute board is based on a Rockchip SoC that exposes a USB ``Loader mode'' allowing direct access to the device’s flash storage. As shown in \autoref{fig:dump}, by connecting over USB and invoking the public rkdeveloptool utility, an attacker can enumerate and dump all GPT partitions including \texttt{uboot}, \texttt{boot}, \texttt{recovery}, \texttt{rootfs}, and \texttt{userdata}, without any authentication, debug authorization, or secure-boot enforcement. The absence of access control on this interface enables full firmware extraction, revealing embedded credentials, configuration files, control binaries, and cryptographic material used throughout the robot’s software stack. Moreover, Sce partition writes are also permitted, an attacker can inject modified system images or persistent backdoors, bypasSg all higher-layer Bluetooth, Wi-Fi, or cloud-level security guarantees. This renders the physical device susceptible to permanent compromise and makes it impossible for a legitimate owner to verify the integrity of the robot’s software state.

\section{Discussion}
\label{sec:discuss}

\subsection{Lessons Learned}

Our analysis of the Unitree Go2 reveals a broader set of lessons for the emerging field of embodied-intelligence security. First, security weaknesses in embodied systems are fundamentally cross-layer: seemingly small flaws in wireless provisioning or certificate handling can cascade into privilege escalation, remote command injection, or persistent firmware compromise. Traditional assumptions from mobile and IoT security fail to hold when conversational interfaces, sensor feedback loops, locomotion control, and cloud-based ownership semantics are tightly coupled. Second, the intelligence layer alone, particularly LLM-centric jailbreak defenses, is insufficient. Even a perfectly aligned AI model cannot compensate for fragile cryptography, weak authentication, or unprotected debug pathways. Effective defenses must therefore begin with a robust physical and firmware trust anchor, enforce authenticated communication across all provisioning and control channels, and isolate safety-critical actuation pathways from higher-level or user-facing logic. Third, embodied systems require a new paradigm of “secure-by-default autonomy”: designs must assume adversarial environments, malicious network conditions, multi-language manipulation, and untrusted local devices as part of the threat model. Finally, the findings underscore the importance of transparency and verifiable system behavior. Without auditable firmware, accountable cloud-binding workflows, and predictable safety boundaries, developers and operators cannot reason about failure modes or detect compromise. As embodied intelligence becomes increasingly deployed in homes, workplaces, and public environments, these lessons highlight the urgent need for cross-disciplinary security principles that unify robotics, networking, AI safety, and systems security into a coherent defense strategy.

\subsection{Mitigation and Defense} 

Securing embodied-intelligence platforms such as the Unitree Go2 requires mitigation strategies that transcend ad-hoc patches and instead address the systemic weaknesses revealed across the three architectural layers of the robot. At the provisioning layer, the most critical step is eliminating fixed cryptographic material by enforcing per-device, hardware-rooted key derivation and mutually authenticated onboarding. BLE and Wi-Fi setup must adopt ephemeral session keys, challenge–response protocols, and strict certificate validation to prevent passive sniffing, replay, and man-in-the-middle interception. For the core software stack, the platform needs a hardened trust boundary: the local HTTP relay should be removed or require authenticated IPC; WebView and JS bridge interfaces must adopt strict origin checking, capability isolation, and signature-based command authorization; and the cloud binding workflow must include device-bound secrets, cryptographic proofs of ownership, and irreversible hardware-level identity. AI-module safety inconsistencies should be addressed through multilingual alignment, guardrail unification, and decoupling of conversational interfaces from physical actuation pathways to prevent jailbreak-to-motion escalation. Finally, the most consequential mitigations lie in the low-level firmware layer: the Rockchip Loader Mode must be locked down using secure-boot verification, authenticated flashing, debug-port gating, and anti-rollback protections. Without these foundational guarantees, any upper-layer security improvement can be trivially bypassed by persistent firmware tampering. End-to-end protection for embodied-intelligence systems therefore requires a holistic redesign of trusted computing bases, key-management lifecycles, remote-ownership semantics, and human–robot interaction boundaries. Only with such cross-layer hardening can embodied robots safely operate in open, adversarial, or human-centric environments.

\section{Related Work}
Research on Embodied AI safety has rapidly expanded in recent years as embodied agents such as autonomous robots, interactive mobile systems, and multimodal decision-making agents—are increasingly deployed in human-centered environments. Early survey efforts, such as \cite{xing2025towards}, provide a comprehensive overview of the vulnerabilities and attack surfaces unique to EAI. These works highlight that EAI systems differ significantly from conventional machine learning due to the tight coupling between perception, language grounding, planning, and physical actuation. As a consequence, unsafe model behavior can directly translate into physical harm, motivating an urgent need for systematic safety analysis.

\hspace{4mm} A rapidly growing line of research focuses on jailbreak attacks against embodied agents. Several works demonstrate that LLM-driven robots can be manipulated through natural-language interactions in black-box settings. For example, \cite{robey2025jailbreaking,lu2024poex} show that carefully crafted prompts or policy-executable instructions can cause robots to violate safety constraints, even without access to internal model parameters. Beyond attacks, recent studies explore defenses: \cite{yang2025concept} introduces concept-enhancement–based robust prompting, while \cite{zhang2024badrobot} extends jailbreak benchmarks to physical-world scenarios. Together, these works reveal that natural-language interfaces significantly broaden the attack surface of embodied systems and that reliable defenses remain highly challenging. In parallel, extensive research investigates adversarial attacks and robustness in multimodal or embodied perception–action loops. Prior studies demonstrate that LLM/VLM/VLA-based robotic controllers are vulnerable to multimodal perturbations, including adversarial images, misleading environmental cues, and corrupted action-state observations. Representative examples include \cite{wu2024highlighting}, which exposes safety risks when deploying LLM/VLM models for robotic control, and \cite{wang2025exploring}, which studies adversarial vulnerabilities of VLA models. Further work examines adversarial impacts on quadrupedal locomotion \cite{shi2024rethinking}, decision-level attacks \cite{liu2024exploring}, navigation-path manipulation \cite{islam2024malicious}, and embodied red-teaming for robotic foundation models \cite{karnik2024embodied}. These findings collectively demonstrate that physical and multimodal adversarial robustness remains a core open problem for EAI.

\hspace{4mm} A separate but related line of work investigates backdoor attacks in embodied systems. Studies such as \cite{wangtrojanrobot,liu2024compromising,jiao2024can,zhou2025badvla} show that malicious triggers injected into training data, prompt contexts, or task specifications can cause embodied agents to execute dangerous or unintended actions while remaining stealthy in benign settings. These backdoor mechanisms pose significant risks due to the complexity of multimodal inputs and the difficulty of detecting distributed triggers in embodied pipelines.
Beyond explicit security attacks, researchers have also explored prompt-injection vulnerabilities. For instance, \cite{zhang2024study} demonstrates that LLM-integrated mobile robotic systems can be coerced into harmful behaviors through carefully crafted prompt-injection patterns, raising concerns about autonomous task execution in open environments. 
Finally, several works examine alignment, safety frameworks, and sociotechnical risks. Studies such as \cite{hundt2025llm} investigate harmful social biases in robot behaviors, while evaluation frameworks like EARBench \cite{zhu2024earbench} assess physical risk awareness in task planning. Recent work also explores safety alignment for VLA models \cite{zhang2025safevla}, responsible manipulation \cite{ni2024don}, and system-level safety frameworks \cite{zhang2024safeembodai}. These efforts emphasize that EAI safety requires not only robust algorithms but also integrated frameworks and governance mechanisms for reliable real-world deployment.

\hspace{4mm} Although prior work has extensively examined jailbreaks, multimodal adversarial inputs, and backdoor risks in embodied AI, these efforts remain largely focused on the model itself. Our study shows that real-world embodied systems face a much broader set of vulnerabilities rooted in provisioning protocols, network interfaces, mobile control paths, cloud binding, and firmware access. Unlike model-centric analyses, we provide the first full-stack audit of a widely deployed quadruped robot and uncover ten systemic flaws that arise from weaknesses across the entire software–hardware pipeline. This system-level perspective reveals practical attack surfaces that existing model-level research does not capture.
\smallskip

\section{Conclusion}
\label{sec:conclusion}

\noindent
We presented the first holistic security assessment of a widely deployed embodied-intelligence platform. Through analyzing the Unitree Go2 across provisioning, networking, cloud workflows, local interfaces, and hardware access, we identified ten systemic vulnerabilities that together form a cross-layer ``Ten Sins'' threatening device integrity, user safety, and physical control.
Our findings show that securing embodied AI requires more than aligning the model. Weak cryptography, unsafe onboarding paths, inconsistent multilingual safeguards, fragile binding logic, and unprotected firmware surfaces can all undermine the system regardless of model behavior. Strengthening embodied AI therefore demands a whole-system security perspective that integrates robust cryptography, authenticated control channels, hardened ownership semantics, and locked-down hardware roots of trust.

\section*{Acknowledgement}

This research was supported by the Quancheng Laboratory Award (QCL20250106). Any opinions, findings, conclusions, or recommendations expressed are those of the authors and not necessarily of the Quancheng Laboratory.

\bibliographystyle{IEEEtranS}
\bibliography{cas-refs}

\end{document}